\documentclass[prd,nofootinbib,floats,aps,twocolumn,tightenlines,superscriptaddress,floatfix]{revtex4-1}

\usepackage{epsf,epsfig,amsmath,amssymb,dsfont}
\usepackage{amsthm,mathrsfs} 
\usepackage{graphicx} 
\usepackage{multirow}
\usepackage{float}
\usepackage{colortbl}
\usepackage[table]{xcolor}
\usepackage{varioref}
\usepackage{slashed}
\usepackage{float}
\usepackage{bm}
\usepackage{hyperref}
\usepackage{longtable}
\usepackage{array}
\usepackage{mathtools}
\usepackage{subfigure}

\newcommand{\ket}[1]{{\left| {#1} \right>}}
\newcommand{\bra}[1]{{\left< {#1} \right|}}

\newcommand{\ii}{\mathrm{i}}

\newcommand{\be}{\begin{equation}}
\newcommand{\ee}{\end{equation}}
\newcommand{\baln}{\begin{align}}
\newcommand{\ealn}{\end{align}}
\newcommand{\ben}{\begin{equation*}}
\newcommand{\een}{\end{equation*}}
\renewcommand{\d}{\mathrm{d}}

\long\def\symbolfootnote[#1]#2{\begingroup%
\def\thefootnote{\fnsymbol{footnote}}\footnote[#1]{#2}\endgroup}

\newcommand{\fr}{\frac}

\begin{document}

\title{
Transmission of Information in Non-Local Field Theories}

\author{Alessio Belenchia}
\affiliation{Institute for Quantum Optics and Quantum Information (IQOQI), Boltzmanngasse 3 1090 Vienna, Austria.}
\author{Dionigi M. T. Benincasa}
\affiliation{SISSA - International School for Advanced Studies, Via Bonomea 265, 34136 Trieste, Italy.}
\affiliation{INFN, Sezione di Trieste, Trieste, Italy.}
\author{Stefano Liberati}
\affiliation{SISSA - International School for Advanced Studies, Via Bonomea 265, 34136 Trieste, Italy.}
\affiliation{INFN, Sezione di Trieste, Trieste, Italy.}
\author{Eduardo Mart\'{i}n-Mart\'{i}nez}
\affiliation{Institute for Quantum Computing, University of Waterloo, Waterloo, Ontario, N2L 3G1, Canada}
\affiliation{Department of Applied Mathematics, University of Waterloo, Waterloo, Ontario, N2L 3G1, Canada}
\affiliation{Perimeter Institute for Theoretical Physics, 31 Caroline St N, Waterloo, Ontario, N2L 2Y5, Canada}


\begin{abstract}

The signaling between two observers in 3+1 dimensional flat spacetime coupled locally to a non-local field is considered. We show that in the case where two observers are purely timelike related -- so that an exchange of on-shell massless quanta cannot occur -- signaling is still possible because of a violation of Huygens' principle. In particular, we show that the signaling is exponentially suppressed by the non-locality scale. Furthermore, we consider the case in which the two observers are light-like related and show that the non-local modification to the local result is polynomially suppressed in the non-locality scale. This may have implications for phenomenological tests of non-local theories. 
\end{abstract}

\maketitle
\flushbottom

\section{INTRODUCTION}\label{secI} 
Assuming standard local dynamics, the geometry of spacetime fully determines how free particles propagate. In particular, the strong Huygens' principle (SHP) states that for flat spacetimes of even dimensions, $d > 2$, the Green function of a free massless theory only has support on the lightcone, that is, information propagates only along null geodesics. In the presence of curvature however (or for flat spacetimes in an odd number of spacetime dimensions) the SHP for massless particles ceases to hold in general \cite{Ellis,McLenaghan,Sonego:1991sq,czapor}, so that information can in principle be transmitted inside the light cone, similarly to what happens with free massive particles in $d=4$.
Violation of the SHP implies that two observers, Alice (the sender) and Bob (the receiver), both locally coupled to a massless quantum field, are able to communicate information via the  field without necessarily having Alice send real, on-shell quanta to Bob~\cite{Jonsson2015}. In fact, it was further shown in~\cite{Jonsson2015} that this timelike information can be accessed by Bob only if he is willing to pay for it (by spending energy), and that no energy is directly transmitted from Alice to Bob, as would be expected given that no on-shell quanta was exchanged. For this reason this process was dubbed ``quantum collect calling''.



Nonlocal field theories provide another example of theories for which the SHP ceases to hold even in $d=4$ flat spacetime. In particular, we have in mind a class of free, massless, nonlocal scalar field theories whose dynamics are defined by $f(\Box)\phi=0$,
where $f$ is a non-analytic function with a branch cut for timelike momenta $p^2\le0$, 
and which exhibit generic violations of the SHP  \cite{Belenchia:2014fda}. 
Theories of this kind arise naturally in the context of causal set theory~\cite{PhysRevLett.59.521} -- an approach to quantum gravity where spacetime is taken to be fundamentally discrete -- where the interplay 
between discreteness and Lorentz invariance inevitably leads to a nonlocality of the type $f(\Box)$ \cite{Sorkin2006,Belenchia:2014fda,Belenchia:2015hca,Benincasa:2010ac,Dowker:2013vba,Glaser:2013sf,BelenchiaUni}.


In this work we investigate the signaling between two observers, Alice (the sender) and Bob (the receiver), living in a flat 4-dimensional spacetime, locally coupled to a non-local, massless scalar field via two-level Unruh-DeWitt particle detectors. 
Since the nonlocal field violates the SHP
one might expect a non-vanishing signaling contribution between the two observers even when they are purely timelike separated, something that would be impossible (in the absence of reflecting mirrors) if the field satisfied the usual local dynamics. Furthermore, one might also expect a modification to the signaling when the observers are in light-like contact, since the response of a single Unruh--DeWitt detector alone is already modified by the non-locality significantly \cite{PhysRevD.94.061902}. As we will show, in the timelike case the signaling is exponentially suppressed in the non-locality scale, whereas a polynomial suppression is present in the case of light-like contact between the detectors. 
While the latter case opens an interesting window for constraining the non-locality scale, the former indicates that Huygens violations are sensitive to the UV structure of the theory. 



The paper is organized as follows. In section~\ref{secII} we review how signaling between two observers is computed in perturbation theory. In section~\ref{secIII}, we introduce the non-local field theory under investigation and, in particular, its Pauli--Jordan function (expectation of the field commutator), which is the central object of this work. In section~\ref{secIV} we compute the non-local contribution to the signaling between Alice and Bob in the cases where they are timelike related and lightlike related. In section~\ref{secV} we discuss the findings of our work and in section~\ref{secVI} we conclude with a summary and outlook. 
 
\section{Setup}
\label{secII}

We will consider a setup consisting of two partners. A sender (Alice) and a receiver (Bob) operating quantum antennas modeled as Unruh--DeWitt particle detectors \cite{DeWitt1980}. 
As well as being very simple, these particle detector models also have the advantage of capturing the fundamental features of light-matter interaction in scenarios where the exchange of angular momentum does not play a prominent role \cite{Martin-Martinez2013,Alhambra2014,Pozas2016}.

This signalling setup has been studied in previous works regarding the exchange of information in local quantum field theory \cite{Jonsson2014,Jonsson2015,Bounce2,Bounce12,Blascoprd,Blasco:2015eya,Martin-Martinez2015,Jonsson2014,Landulfo2016,Petar,Jonsson2015,Jonsson1,Jonsson2,Petar2017}.

The Unruh--DeWitt interaction Hamiltonian coupling an inertial particle detector to a scalar field is given by
\begin{equation}
   \label{puno} H_{I,\nu}=\lambda_\nu \chi_\nu(t) m_\nu(t) \phi[\bm x(t),t],
\end{equation}
where $\nu\in\{\text{A,B}\}$ is a label indicating Alice's and Bob's detectors, $\lambda_\nu$ are coupling constants, $\chi_\nu(t)$ are the switching functions controlling the coupling-decoupling speed and the duration of the detector-field interaction, \mbox{$m_\nu(t) =(\sigma_\nu^+e^{\ii\Omega_\nu t}+\sigma_\nu^-e^{-\ii\Omega_\nu t})$} are the detector's monopole moment ($\Omega_\nu$ is the energy gap between detector $\nu$'s two energy levels), and $\phi[\bm x_\nu(t),t]$ is the field evaluated along the detector's trajectories.
In the following we set $c=1$.

We are going to consider the same signalling setup as in \cite{Jonsson2015,PhysRevD.93.024055,Blasco:2015eya,Petar}. Let each detector start out in the pure state \mbox{$\rho_{0,\nu}=\ket{\psi_{0,\nu}} \bra{\psi_{0,\nu}}$}, where $\ket{\psi_{0,\nu}}=\alpha_\nu\ket{e_\nu}+\beta_\nu\ket{g_\nu}$ with $\ket{g(e)_{\nu}}$ the ground state and excited state of detector $\nu$ respectively, and let the field start out in an arbitrary state $\rho_{0,\phi}$. Hence, the initial state of the system is
	\begin{equation}
	\rho_0=\rho_{0,\textsc{a}} \otimes \rho_{0,\textsc{b}} \otimes \rho_{0,\phi}.
	\end{equation}
	Allowing the system to evolve under the full interaction Hamiltonian $H_\textsc{i}(t)=H_{\textsc{i},\textsc{a}}(t)+H_{\textsc{i},\textsc{b}}(t)$ for a time $T$ results in the state $\rho_{_T}=U\rho_0 U^\dagger$, where U is the time evolution operator
	\begin{equation}
	    U=\mathcal{T}\exp\left[-\ii\int_{-\infty}^\infty \!\!\d t H_\textsc{i}(t)\right],
	\end{equation}
	and $\mathcal{T}$ denotes time-ordering of the exponential. The final state of Bob's detector is obtained by tracing out the field and the state of Alice from $\rho_{_T}$:
	\begin{equation}\label{eq:rho_t_b}
	\rho_{_{T,\text{B}}}=\text{tr}_{\phi,\textsc{a}}(\rho_{_T}).
	\end{equation}

Consider the situation in which Alice --- the sender --- turns on her detector for a time $T$ and, subsequently, Bob --- the receiver --- turns on his own detector\footnote{Here we will consider the case in which Alice and Bob are at rest relative to each other.}. The probability that Bob (who is in the causal future of Alice) finds his detector in the excited state after a time $T^*>T$ from when he turns it on can be written as
\begin{equation}\label{probBob}
P_{e}(T^*)=|\alpha|^2+R(T^*)+S(T^*).
\end{equation}
The first term in \eqref{probBob} stands for the probability of the initial state to be excited, $R(T^*)$ is noise due to the coupling of Bob's detector to the field for a finite time, and the last term is the signaling term that depends on Alice being coupled to the field in the past.

At leading order in perturbation theory the signaling term can be written as $S=\lambda_\textsc{a}\lambda_\textsc{b}S_2+\mathcal{O}(\lambda_\nu^4)$ where $S_{2}$ is given by (see \cite{Jonsson2015})
\begin{align}\label{signal}
S_{2}=&4\lambda_{\textsc{b}}\lambda_{\textsc{a}}\int \rm{d} t_{2}\int \rm{d}t_{1}\chi_{\textsc{a}}(t_{1})\chi_{\textsc{b}}(t_{2})Re(\alpha_{\textsc{a}}^* \beta_{\textsc{a}}e^{i\Omega_{\textsc{a}}t_{1}})\\ \nonumber
& Re(\alpha_{\textsc{b}}^*\beta_{\textsc{b}}e^{i\Omega_{\textsc{b}}t_{2}}[\phi(x_{\textsc{a}},t_{1}),\phi(x_{\textsc{b}},t_{2})]),
\end{align} 
the $\chi$s are the detectors' switching functions\footnote{We assume $\rm{supp}(\chi_{A})\cup \rm{supp}(\chi_{B})=\emptyset$.} and the integration is over the times $t_{1(2)}$ of Alice and Bob respectively.

The last expression contains the Pauli--Jordan function of the field theory. This guarantees that, if Huygens' principle holds, the signaling between timelike related detectors vanishes. Moreover, in a causal theory\footnote{A theory which respects (micro)-causality, i.e., $[\phi(\mathsf{x}),\phi(\mathsf{y})]=0$ $\forall\;\;\mathsf{x},\mathsf{y}$ spacelike.}, the signaling between spacelike observers also vanishes as well so that no faster-than-light signaling is allowed.  

The signalling term allows for communication of Alice with Bob, even if they are timelike separated, as long as the commutator between the spacetime regions where Alice and Bob exist is non-vanishing. Lower bounds to the channel capacity (in bits per use of the channel) have been studied by setting up a concrete communication protocol in several scenarios in flat \cite{Jonsson2015,Robert1} and curved spacetimes \cite{PhysRevD.93.024055,Blasco:2015eya,Petar}. Notice that protocols that optimize the choice of the initial state of Alice's detector and the measured observable on Bob's detector have also been considered in previous literature \cite{Jonsson2}. 

In 4D flat spacetime it is not possible to communicate through a local massless scalar field, when the parties are purely timelike related, since the commutator is non-zero only between light-connected events. However non-localities in the theory will induce modifications of the field commutator that will manifest in two ways: they will enable some form of timelike communication and they will modify the channel capacity through lightlike communication.

\section{NON-LOCAL FIELD THEORY: PAULI--JORDAN FUNCTION}\label{secIII}

We are interested in a massless scalar field
with dynamics given by a real, retarded, Poincar\'e invariant operator, $\widetilde{\Box}\coloneqq f(\Box)$. It can be shown \cite{Aslanbeigi:2014zva} that operators of this kind have Fourier transforms which depend both on $k^2$ -- as one would expect from a Lorentz invariant operator --  and $\text{sgn}(k^0)$; the latter property being a consequence of the retarded nature of these operators. That is
\be
\widetilde{\Box}e^{\ii k^\mu  x_\mu}=B(\text{sgn}(k^0),k^2)e^{\ii k^\mu  x_\mu}.
\ee
The dependence on $\text{sgn}(k^0)$ implies that the function $B$ possesses a branch cut along $k^2\le 0$ that is associated to a continuum of massive modes.

It is possible to construct quantum field theories based on these operators. In particular, it can be shown that the Wightman two-point function of these theories is given by
\begin{align}
D^{(+)}(x-y) = &\int \fr{\rm{d}^4k}{(2\pi)^4} \widetilde{W}(k^2) e^{\ii \mathsf{k}\cdot(\mathsf{x}-\mathsf{y})},
\label{wightman}
\end{align}
with
\be
\widetilde{W}(k^2) = \fr{2\text{Im}(B)\theta(k^0)}{|B|^2}.
\label{discont} 
\ee
$\widetilde{W}$ can be rewritten as $\widetilde{W}(k^2)=2\pi \widetilde{\rho}(-k^2)$ where $\widetilde{\rho}$ can be further split into a divergent part and a finite part: $\widetilde{\rho}(\mu^2)=\delta(\mu^2)+\rho(\mu^2)$.
With this is mind one can see that $D^{(+)}$ can actually be written as a sum of two parts:  one being the standard Wightman function for a local massless scalar field, $D^{(+)}_0$, and the other an integral over the Wightman function of a local massive field, $G^{(+)}_\mu$, weighted by the finite part of the discontinuity function, $\rho(\mu^2)$, i.e.
\begin{align}\label{wight}
&D^{(+)}(x-y) = \int \fr{\rm{d}^4k}{(2\pi)^4} 2\pi\theta(k^0)\delta(k^2)e^{\ii \mathsf{k}\cdot(\mathsf{x}-\mathsf{y})} \\
&+\int_0^\infty \rm{d}\mu^2 \rho(\mu^2)\int \fr{\rm{d}^4k}{(2\pi)^4} 2\pi\theta(k^0)\delta(k^2+\mu^2)e^{\ii \mathsf{k}\cdot(\mathsf{x}-\mathsf{y})}. \nonumber
\end{align}

For every choice of $\widetilde{\Box}$ there corresponds a specific $\rho$. In this paper we are interested in a discontinuity function of the form
\be\label{2disc}
\rho(\mu^2)
=\ell^2e^{-\alpha \ell^2\mu^2}.
\ee
where $\alpha$ is an order one numerical coefficient~\cite{Saravani:2015rva}. This choice of $\rho$ is a simple function which captures all the fundamental features of more complex spectral densities  in previous literature (see~\cite{Belenchia:2014fda,PhysRevD.92.103504} and references therein).

\subsection{Pauli-Jordan}
Generalizing the previous discussion on the Wightman function to the Pauli-Jordan function it should be clear that the latter will also
decompose into a sum of a local massless Pauli-Jordan function and the integral over all masses of massive Pauli-Jordan functions weighted by the spectral density, i.e.
\begin{equation}
[\phi(\mathsf{x}),\phi(\mathsf{y})]=[\phi(\mathsf{x}),\phi(\mathsf{y})]_{0}+\int \rm{d}\mu^{2}\rho(\mu^{2}) [\phi(\mathsf{x}),\phi(\mathsf{y})]_{\mu},
\end{equation}
where $[\phi(\mathsf{x}),\phi(\mathsf{y})]_{\mu}$ represents the Pauli-Jordan for a field of mass $\mu$ and is given by~\cite{bogoliubov1959introduction}
\begin{equation}
[\phi(\mathsf{x}),\phi(\mathsf{y})]_{\mu}=[\phi(\mathsf{x}),\phi(\mathsf{y})]_{0}-\frac{\mu}{4\pi\sqrt{-\sigma}}\Theta(-\sigma)J_{1}(\mu\sqrt{-\sigma}),
\end{equation} 
with $\sigma=-\Delta t^2+\Delta x^2$ and $J_{1}$ is a Bessel function of the first kind. Thus, we can rewrite the commutator of the non-local field as
\begin{align}\label{PJ}
[\phi(\mathsf{x}),\phi(\mathsf{y})]&=[\phi(\mathsf{x}),\phi(\mathsf{y})]_{0}+\frac{1}{\alpha}[\phi(\mathsf{x}),\phi(\mathsf{y})]_{0}\\ \nonumber
&+\int \rm{d}\mu^{2}\rho(\mu^{2})\left(-\frac{\mu}{4\pi\sqrt{-\sigma}}\Theta(-\sigma)J_{1}(\mu\sqrt{-\sigma})\right),
\end{align}
where the $1/\alpha$ in the second term is given by the integral over all $\mu$ of \eqref{2disc}. From now on we will set $\alpha=1$ for convenience of notation. For $\sigma<0$, the last term on the RHS of \eqref{PJ} is given by 
\begin{equation}
-\frac{1}{8\pi \ell^2}e^{\frac{\sigma}{4\ell^{2}}},
\end{equation}
so that the commutator takes the following form
\begin{equation}
[\phi(\mathsf{x}),\phi(\mathsf{y})]=2[\phi(\mathsf{x}),\phi(\mathsf{y})]_{0}-\frac{1}{8\pi \ell^2}e^{\frac{\sigma}{4\ell^{2}}}\Theta(-\sigma).
\end{equation}

\subsection{Distributional local limit}
At first sight the above expression may seem puzzling because the local result appears to have been modified by an $\ell$-independent term (having chosen $\alpha=1$, a factor of two appears in front of the local result). However, this is just an artefact. 
Indeed, the quantity
\begin{equation}\label{dw}
-\frac{1}{8\pi \ell^2}e^{\frac{\sigma}{4\ell^{2}}}
\end{equation}
converges, in a weak sense, to $-\delta(\sigma)/2\pi$ for vanishing $\ell$. 
This can be checked explicitly by integrating the above expression against test functions and then taking the local limit --- we will see this in the next sections. Thus, in the local limit the Pauli--Jordan function converges weakly to the local result --- the second term in \eqref{PJ} being cancelled by the weak limit of the last one.

\section{NON-LOCAL SIGNALING CONTRIBUTION}\label{secIV}
Given the splitting of the Pauli-Jordan function we can write $S_{2}$ in \eqref{signal} as
\begin{equation}\label{S2}
S_{2}=S_{2}^{local}+S_{2}^{non-local},
\end{equation} 
where $S_{2}^{non-local}=S_{2}^{local}+S_{2}^{(\ell)}$ and the second term is the one that is  dependent on the non-locality scale $\ell$. The local limit is recovered since  $S_{2}^{(\ell)}$ converges to $-S_{2}^{local}$ in the $\ell\rightarrow 0$ limit. 

Without loss of generality we will assume that $\alpha_{X}$ and $\beta_{X}$ are real for $X=A,B$, and also that $\Omega_{\textsc{a}}=\Omega_{\textsc{b}}=\Omega$. The $\ell$ dependent term in \eqref{S2} then becomes
\begin{align}
S_{2}^{(\ell)}&=-4\alpha_{\textsc{b}}\beta_{\textsc{b}}\alpha_{\textsc{a}}\beta_{\textsc{a}}\int \rm{d}t_{2}\chi_{\textsc{b}}(t_2)\int  \rm{d}t_{1}\chi_{\textsc{a}}(t_1)\\ \nonumber
&\times\cos(\Omega t_{1})\cos(\Omega t_{2})\frac{1}{8\pi\ell^2}e^{\frac{-\Delta t^2+R^2}{4\ell^2}}\Theta(-\sigma).
\end{align}
We consider different configurations of the two detectors for various switching functions.

\subsection{Bob in the lightband of Alice: polynomial suppression}
\begin{figure*}
\includegraphics[width=0.45\textwidth]{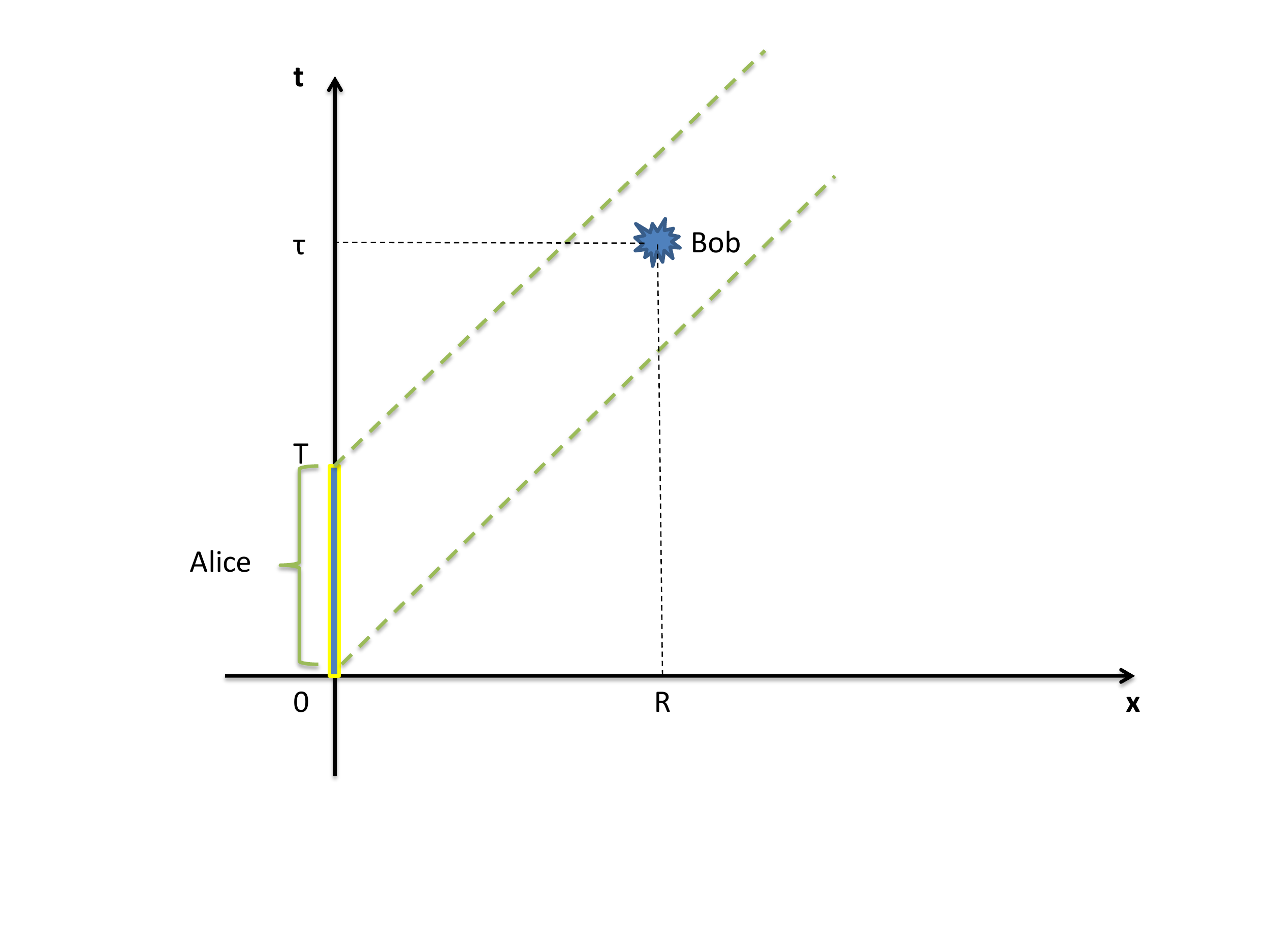}\quad
\includegraphics[width=0.45\textwidth]{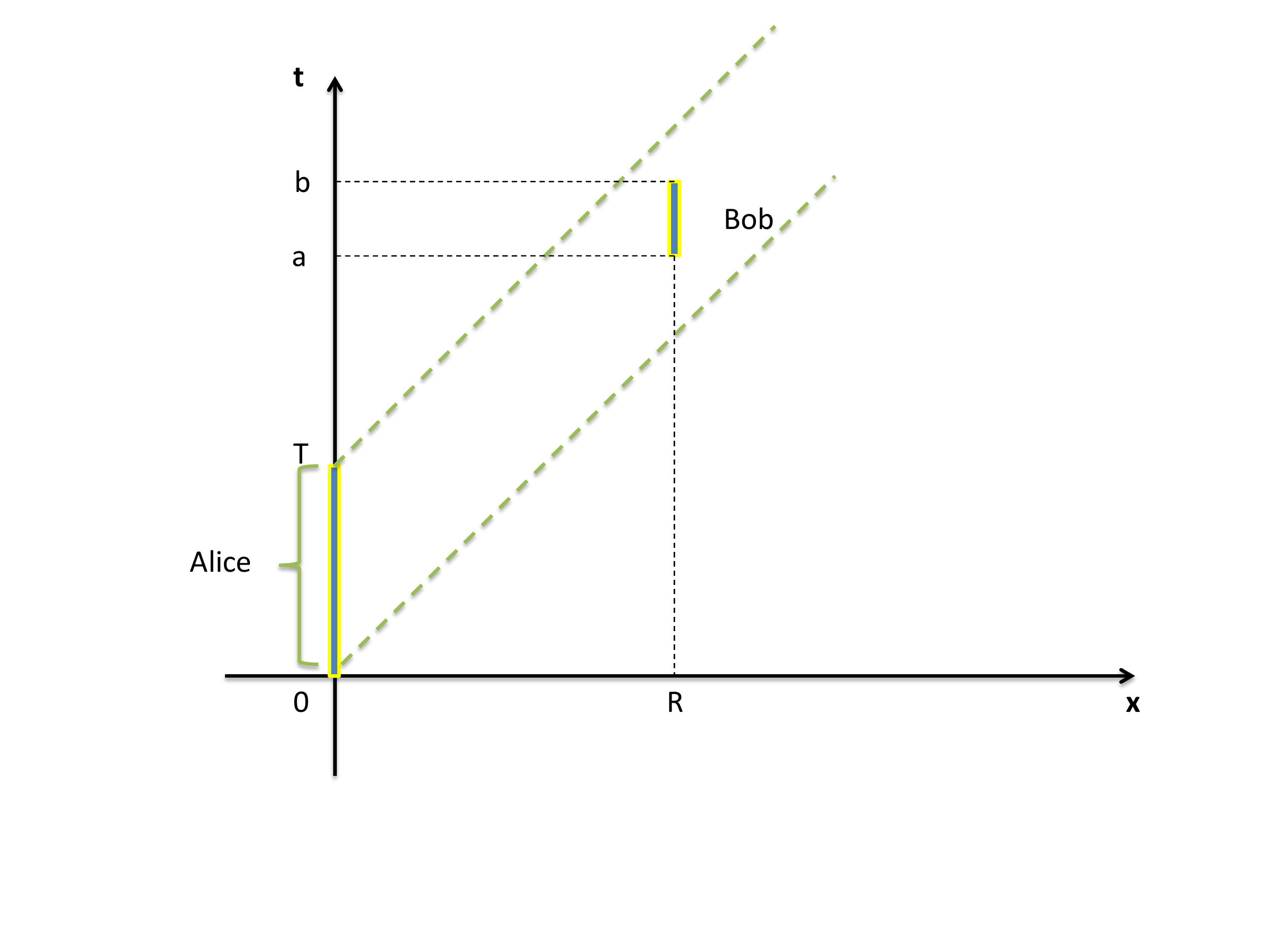}\quad
  \caption{Left, spacetime diagram sketching the configuration of the two detectors in which Bob is in the ``lightband'' of Alice and has a delta-switching detector. Right, same configuration as before with Bob detector ``on'' for a finite amount of time.}
  \label{fig1}
\end{figure*}

We start by investigating the case in which Bob is inside the lightband of Alice, see Fig.\ref{fig1}. In this case the two observers are in lightlike contact and can exchange real quanta of a massless field.  Signaling in this configuration is therefore allowed both in the local and non-local theory, and we are interested in how the non-locality modifies the  non-vanishing $S_{2}$ computed from a local theory.

As our calculations will show, the nonlocal correction to the signaling between Alice and Bob in this case is polynomially suppressed in the nonlocality scale. Furthermore, we will argue that the polynomial suppression does not depend on the UV details of the discontinuity function. 


\subsubsection{Case 1: delta-switching}
Consider the case in which Alice's detector is suddenly switched on at 0 and then suddenly back off at $T$ (note that while this would introduce divergences in the detector's response in 3+1D that would have to be regularized, as discussed in \cite{Jonsson2015,Blasco:2015eya}, the signalling term is devoid of any UV divergences), whereas Bob's detector is on only at $t=\tau$, with $\tau>T$. We also assume that $R<\tau<R+T$, where $R$ is the constant spatial distance between the two observers. This is tantamount to assuming Bob is inside Alice's light band (see the left panel in Fig.\ref{fig1}). In this case 
\begin{align}
\mathcal{S}_{2}^{(\ell)}&\propto \int \rm{d}t_{2}\int_{0}^{T}\rm{d}t_{1}\cos(\Omega t_1)\cos(\Omega t_2)\frac{1}{8\pi \ell^{2}}e^{\frac{-(t_1-t_2)^2+R^2}{4\ell^2}}\\ \nonumber
& \times\Theta((t_2-t_1)^2-R^2)\delta(t_2-\tau)\\ \nonumber
&= \int_{0}^{T}\rm{d}t_{1}\cos(\Omega t_1)\cos(\Omega \tau)\frac{1}{8\pi \ell^{2}}e^{\frac{-(t_1-\tau)^2+R^2}{4\ell^2}}\\ \nonumber
&\times\Theta((\tau-t_1)^2-R^2)\\ \nonumber
&= \int_{0}^{\tau-R}\rm{d}t_{1}\cos(\Omega t_1)\cos(\Omega \tau)\frac{1}{8\pi \ell^{2}}e^{\frac{-(t_1-\tau)^2+R^2}{4\ell^2}}.
\end{align}
The integral above can be computed analytically and the result is given by
\begin{align}
\mathcal{S}_{2}^{(\ell)}&=-4\alpha_{\textsc{b}}\beta_{\textsc{b}}\alpha_{\textsc{a}}\beta_{\textsc{a}}\frac{\kappa}{16 \sqrt{\pi } \ell}\left[\cos (\tau  \Omega ) e^{\frac{R^2}{4 \ell^2}-\Omega  \left(\ell^2 \Omega +\ii \tau \right)}\right.\\ \nonumber
&\times\left(\text{erf}\left(\frac{\tau }{2 \ell}-\ii \ell \Omega \right)+\text{erfc}\left(\frac{R}{2 \ell}-\ii \ell \Omega \right)-1\right.\\ \nonumber
&\left.\left.+e^{2 \ii \tau  \Omega } \left(\text{erf}\left(\frac{\tau }{2 \ell}+\ii \ell \Omega \right)-\text{erf}\left(\frac{R}{2 \ell}+\ii \ell \Omega \right)\right)\right)\right],
\end{align}
where $\text{erf}$ and $\text{erfc}$ are the error function and complementary error function respectively. Note here that $\kappa$ has dimensions of length (needed to dimensionally balance Bob's delta-function switching) and characterizes how much of an energy perturbation we introduce with the ``kick". 
Consistently with our previous discussion, the local limit of the above expression --- obtained by taking the limit for vanishing $\ell$ --- coincides with minus $S_{2}^{local}$, i.e. the signaling in the local case which is given by
\begin{align}
\mathcal{S}_{2}^{local}&=\frac{2k\alpha_{\textsc{b}}\beta_{\textsc{b}}\alpha_{\textsc{a}}\beta_{\textsc{a}}}{\pi}\int \rm{d}t_{2}\int_{0}^{T}\rm{d}t_{1} \delta((t_2-t_1)^2-R^2)\\ \nonumber
&\times \cos(\Omega t_1)\cos(\Omega t_2)\Theta((t_2-t_1)^2-R^2)\delta(t_2-\tau)\\ \nonumber
&= k\alpha_{\textsc{b}}\beta_{\textsc{b}}\alpha_{\textsc{a}}\beta_{\textsc{a}} \frac{\cos(\Omega \tau)\cos(\Omega\tau-\Omega R)}{\pi R},
\end{align}
where 
$\tau>t_{1}$ by assumption. The fact that $S_{2}^{(\ell)}$ weakly converges to $-S_{2}^{local}$ for vanishing $\ell$ --- and thus Eq.\eqref{S2} gives back the local result --- is consistent with the non-local theory reducing, in the same limit, to the local one. 

In order to determine the leading order correction to the local result when the non-locality scale is small compared to every other scale in the problem, we need to expand the whole non-local contribution to the signaling: $S_{2}^{local}+S_{2}^{(\ell)}$ (see Eq.\eqref{S2}). The leading order correction is given by
\begin{align}\label{corr1}
& k\alpha_{\textsc{b}}\beta_{\textsc{b}}\alpha_{\textsc{a}}\beta_{\textsc{a}}\frac{\ell^2}{\pi R^3}\left[R\Omega\left(\sin(\Omega R)+\sin(R\Omega-2\tau\Omega)\right)\right.\\ \nonumber
&\left.+\cos(\Omega R)+\cos(R\Omega-2\tau\Omega)\right].
\end{align}
This expression shows that the non-local contribution to the signaling is polynomial in the non-locality scale, a fact that resembles the polynomial modification of the response of a single Unruh--DeWitt detector coupled to a non-local field \cite{PhysRevD.94.061902}. 

\subsubsection{Case 2: Bob extended interaction}\label{Irdep}
We now consider the case in which Bob's switching profile is the same as Alice's, that is, sudden switching on and off for a finite amount of time, so that Bob interacts with the field for a finite period of time (see right panel in Fig.\ref{fig1}). We will show that the polynomial suppression of the non-local signaling contribution persists in this case, as one might have expected. 

The non-local signaling contribution with these switching functions is given by
\begin{align}
S_{2}^{(\ell)}=&-4\alpha_{\textsc{b}}\beta_{\textsc{b}}\alpha_{\textsc{a}}\beta_{\textsc{a}}\int_{a}^{b}\rm{d}t_{2}\int_{0}^{T} \rm{d}t_{1}\cos(\Omega t_{1})\cos(\Omega t_{2})\\ \nonumber
&\times\frac{1}{8\pi\ell^2}e^{\frac{-\Delta t^2+R^2}{4\ell^2}}\Theta(-\sigma).
\end{align}
Using the following change of variables 
\begin{align}
& -y\equiv-\Delta t^2+R^2\\
&t_{1}=t_{2}-\sqrt{R^2+y},\\
& dt_{1}=-\frac{1}{2\sqrt{R^2 +y}}dy,
\end{align}
we can rewrite $S_{2}^{(\ell)}$ as
\begin{align}\label{int}
S_{2}^{(\ell)}=&4\alpha_{\textsc{b}}\beta_{\textsc{b}}\alpha_{\textsc{a}}\beta_{\textsc{a}}\int_{a}^{b}\rm{d}t_{2}\cos(\Omega_{\textsc{b}}t_{2})\int_{t_{2}^2 -R^2}^{max[0,(t_2-T)^2-R^2]}\\ \nonumber
&\times\frac{\rm{d}y}{2\sqrt{R^2+y}}\cos(\Omega_{\textsc{a}}(t_{2}-\sqrt{R^2+y})\frac{1}{8\pi\ell^2}e^{\frac{-y}{4\ell^2}}.
\end{align}
Note that we have not yet assumed that Bob is inside Alice's lightband. If the two observers are purely timelike related then $t_{2}>R+T$ and $(t_2-T)^2-R^2>0$, $\forall t_{2}\in [a,b]$, while whenever $t_{2}\leq R+T$ the two detectors will be lightlike related. In the latter case the dependence on $T$ drops out of the signaling because we are only requiring that Bob be inside the lightband of Alice, i.e. $b<T+R$, so that no information on the position of the inner boundary of Alice's lightband is needed, since the signal to Bob comes from Alice's detector from $t=0$ to $t=b-R$. 

\begin{figure}
\includegraphics[width=0.40\textwidth,angle=0]{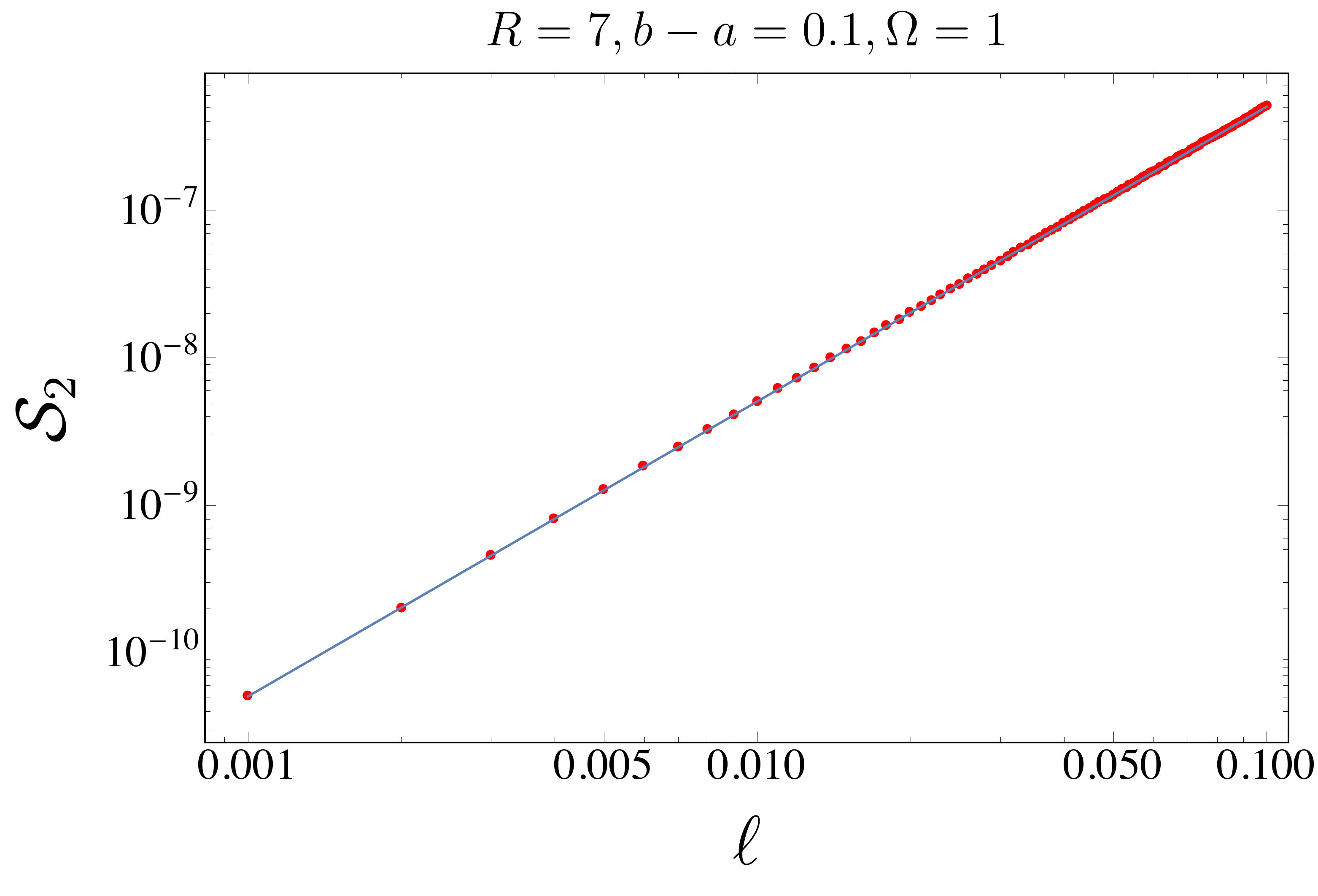}
  \caption{Numerical integration of eq.\eqref{extend} vs. non-locality scale $\ell$. Here we neglected the coupling constants $\lambda_{\nu}$, $\alpha_{\nu},\beta_{\nu}$  as well as multiplicative numerical factors. We chose $R=7$, $a=8$, $b=8.1$, $\Omega=1$. $\ell$ goes from $10^{-3}$ to $10^{-1}$. The (red) points represents the numerical values of $\mathcal{S}_{2}$ whereas the continuum (blue) line is their interpolation with a curve quadratic in $\ell$.
  }
  \label{fig3}
\end{figure}

Since here we are interested in the case in which Bob is inside the lightband of Alice, the expression for $S_{2}^{(\ell)}$ becomes 
\begin{align}\label{extend}
S_{2}^{(\ell)}=&-4\alpha_{\textsc{b}}\beta_{\textsc{b}}\alpha_{\textsc{a}}\beta_{\textsc{a}}\int_{\tilde{a}}^{\tilde{b}}\rm{d}\tilde{t}\cos(\tilde{t})\int_{\tilde{t}^2 -\tilde{R}^2}^{0}\\ \nonumber
&\times\frac{d\tilde{y}}{2\sqrt{\tilde{R}^2+\tilde{y}}}\cos(\tilde{t}-\sqrt{\tilde{R}^2+\tilde{y}})\frac{1}{8\pi\tilde{\ell}^2}e^{\frac{-\tilde{y}}{4\tilde{\ell}^2}},
\end{align}
where we have introduced dimensionless variables --- denoted by tildes\footnote{In the following we drop the tilde for notational convenience.} --- defined in units of $\Omega$, $\tilde{a}\geq R$ and $\tilde{a}<\tilde{b}\leq R+T$. This integral can be computed analytically and the result is given in appendix~\ref{appA}. As before, the local limit of $S_{2}^{(\ell)}$ coincides with $-S_{2}^{local}$. The leading contribution to the signaling coming from $S_{2}^{(\ell)}+S_{2}^{local}$ is again polynomial in $\ell^2$ as is shown in Fig.\ref{fig3}. In the particular case in which $a=R$ and $b=R+T$ this correction (after inserting back dimensional quantities) is given by
\begin{widetext}
\begin{align}\label{lead}
-\alpha_{\textsc{b}}\beta_{\textsc{b}}\alpha_{\textsc{a}}\beta_{\textsc{a}}\ell^{2}\frac{2\Omega^{2} R T \sin (\Omega R)+\sin (\Omega(R+2 T))+\Omega(3 R+2 T) \cos (\Omega R)+\Omega R \cos (\Omega(R+2 T))-\sin (\Omega R)}{2 \pi \Omega R^3}.
\end{align}
\end{widetext}

Finally, in this case it can be also verified (by numerical means) that the polynomial nature of the correction to the local result is independent of the specific UV details of the discontinuity function $\rho$, and indeed holds when one approximates $\rho$ simply by $l^2$, which is exactly what determines the polynomial suppression.

\subsection{Bob timelike to Alice: violations of the SHP}\label{timel}

\begin{figure}
\includegraphics[width=0.45\textwidth]{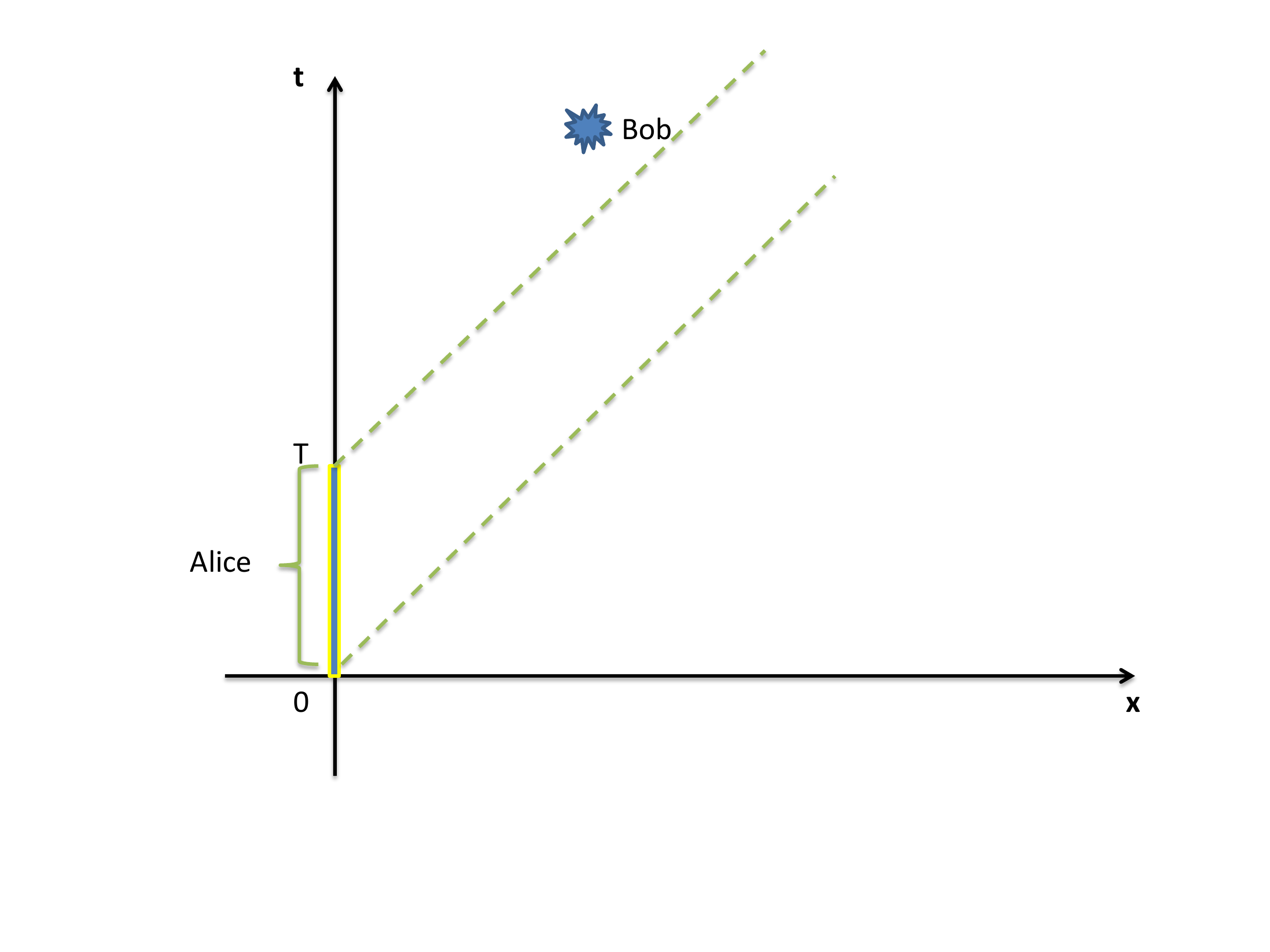}
  \caption{Spacetime diagram sketching the configuration of the two detectors in which Bob is timelike to Alice and has a delta-switching detector.
  }
  \label{fig2}
\end{figure}

In the previous section we investigated the case in which Alice and Bob are allowed to communicate both in the non-local and local theory. In this section we consider the case in which Bob is purely timelike related to Alice, see Fig.\ref{fig2}. 
In a local massless theory (and in the absence of any mirrors) Bob would not be able to receive any information from Alice in this configuration, since no real quanta can be exchanged by the two. 
In the context of a non-local massless theory however, where the SHP fails to hold, one faces the possibility of establishing a communication channel between the two observers, even when they are timelike related. 
From a phenomenological point of view this situation is particularly interesting since it allows for a binary test to asses the presence non-local effects. However, for this to be of any practical interest, we first need to determine how the signaling between Bob and Alice is suppressed by the non-locality scale.

Consider the simple case where Alice's detector is switched on for a finite amount of time, from $0$ to $T$, abruptly, while Bob's detector is on only for an instant of time at $\tau>R+T$, see Fig.\ref{fig2}. In this case $\mathcal{S}$ is given by
\begin{align}
\mathcal{S}_{2}^{(\ell)}&\propto \int \rm{d}t_{2}\int_{0}^{T}\rm{d}t_{1}\cos(\Omega t_1)\cos(\Omega t_2)\frac{1}{8\pi \ell^{2}}e^{\frac{-(t_1-t_2)^2+R^2}{4\ell^2}}\\ \nonumber
&\times\Theta((t_2-t_1)^2-R^2)\delta(t_2-\tau)\\ \nonumber
&= \int_{0}^{T}\rm{d}t_{1}\cos(\Omega t_1)\cos(\Omega \tau)\frac{1}{8\pi \ell^{2}}e^{\frac{-(t_1-\tau)^2+R^2}{4\ell^2}}.
\end{align}
The integral can be computed and gives 
\begin{align}
\label{timelike}
S_{2}^{(\ell)}&=-4\alpha_{\textsc{b}}\beta_{\textsc{b}}\alpha_{\textsc{a}}\beta_{\textsc{a}}\frac{1}{16 \sqrt{\pi } \ell}\cos (\tau  \Omega ) e^{\frac{R^2}{4 \ell^2}-\Omega  \left(\ell^2 \Omega +\ii \tau \right)}\\ \nonumber
&\times\left(\text{erf}\left(\frac{\tau }{2 \ell}-\ii \ell \Omega \right)+\text{erf}\left(\frac{T-\tau }{2 \ell}+\ii \ell \Omega \right)\right.\\ \nonumber
&\left.+e^{2 \ii \tau  \Omega } \left(\text{erf}\left(\frac{\tau }{2 \ell}+\ii \ell \Omega \right)+\text{erf}\left(\frac{T-\tau }{2 \ell}-\ii \ell \Omega \right)\right)\right)
\end{align}
The local limit $\ell\rightarrow0$ can be shown to vanish consistently with the local result of no-signaling.

Note that in the limit where Bob approaches the future boundary of Alice's lightband, i.e., $\tau\rightarrow T+R$, equation \eqref{timelike} becomes of order $l^2$, implying that the polynomial correction to the local result arises from contributions coming from a future-timelike neighbourhood of Alice's future lightband boundary\footnote{Crucially, this result holds true in the limit of vanishing $\ell$ only if $\tau$ approaches $T+R$ faster than the rate at which $\ell$ vanishes.} (in the case where Alice has a delta switching the future lightband boundary reduces to Alice's future lightcone). 

Finally, expanding the result for small $\ell$ shows that in general the suppression of the signaling is exponential in the non-locality scale. Similar results hold if Bob interacts with the field for a finite amount of time. So even if in principle we have a binary test for such non-local phenomena, the fact that the effect is exponentially small makes it of no practical use, unless some amplification effects can be introduced. Note also that, contrary to the lightlike case, 
in this setup the signaling is no longer determined by the IR behaviour of the theory alone (provided the distance from the future boundary of the sender's lightcone is greater than the non-locality scale), since the explicit form
of the suppression depends on the full $\rho$. 


\section{DISCUSSION}\label{secV}
We have shown that when Alice and Bob are lightlike related (meaning that they have the possibility of exchanging real quanta of a massless field) the prediction for the signaling coming from a non-local theory differs from the local one by a correction suppressed only polynomially in the non-locality scale. When Alice and Bob are timelike related, which precludes signaling in a massless, local theory, the correction due to non-locality is exponentially suppressed in $\ell$, unless Bob is in a neighbourhood of Alice's light of cone of size smaller than $O(\ell^2)$.

Given the leading order correction to the signaling when Alice and Bob are lightike related and both interact with the field for a finite amount of time, we can compute the ratio $\mathcal{R}=\left(\mathcal{S}_{2}-\mathcal{S}^{local}_{2}\right)/\mathcal{S}^{local}_{2}$. From eq.\eqref{S2} we know that the ratio is given by 
\begin{equation}
\mathcal{R}=\frac{\mathcal{S}_{2}^{(nl)}}{\mathcal{S}_{2}^{local}},
\end{equation}
which vanishes in the limit of vanishing $\ell$.
The leading order contribution to the above expression is suppressed by $\ell^{2}$. It is interesting to note that the oscillating character of the local result (see Appendix~\ref{appA}), together with the fact that the local result and the non-local correction (eq.~\eqref{lead}) have different zeros in general, allows for an amplification of the non-local signal. Of course, such an amplification requires a fine tuning of the parameters of the system. Another interesting case is the one of degenerate detectors, i.e, vanishing $\Omega$. In this case we get from eq.\eqref{lead} a particularly simple expression for $\mathcal{R}$,
\begin{equation}
\mathcal{R}=8\frac{\ell^2}{R^2}\left(\frac{R+T}{T}\right). 
\end{equation}
In both cases the suppression in the non-locality scale shows that in order to cast stringent constraints on the non-locality scale using signaling between light-like related observers a large sample of events is necessary. This is similar to what happens in the case where the response of a single Unruh--DeWitt detector is used to cast constraints on the non-locality scale, see~\cite{PhysRevD.94.061902}. 

As already discussed in previous sections, this polynomial suppression follows from the IR properties of the spectral function of the non-local theory, i.e. that $\rho\approx\ell^2$ for $\mu^2\rightarrow0$, and is independent of the UV details of $\rho$. To further validate this fact we have numerically computed the signalling for the case in which both Bob and Alice are interacting with the field for a finite amount of time when $\rho$ is set to $\ell^2$. This again gives the expected polynomial suppression in $\ell$.

Regarding the case of time-like related observers, we have shown that the signaling is exponentially suppressed in the non-locality scale --- as far as Bob is bounded away from a neighborhood of Alice's light-cone. Let us stress once again that, the form of the suppression depends on the specific spectral function chosen and cannot be derived just by looking at the IR behaviour of the spectral function. In order to retrieve a polynomial suppression consistent with just the IR property of the spectral function, Bob's detector has to be placed (impractically) close to Alice's lightcone: closer than the non-locality scale we would like to probe. 
This fact, in conjunction with the observation made at the end of section~\ref{Irdep}, that the polynomial correction is only sensitive to the IR behaviour of $\rho$,  strongly suggests that the polynomial correction is not sensitive to the UV also in this scenario. Indeed, since the signal that arises in the case where Bob and Alice are exactly lightlike related does not depend on the nonlocality scale, it is hard to imagine where else the correction may come from if not from a neighbourhood of Alice's lightcone. Ideally one would want to check this explicitly by repeating the calculation of section~\ref{timel} with $\rho=l^2$. Unfortunately we have been unable to perform this computation as of yet, meaning that a definitive claim on this matter is still out of reach.

Finally note that 
the exponential suppression of $\mathcal{S}_{2}$, when the detectors are timelike related, is not a universal feature but a consequence of the particular form of the spectral function Eq.~\eqref{2disc}.\footnote{It is interesting to note that, in causal set theory --- where the same kind of non-local field operators we are discussing were firstly derived from first principles --- the spectral functions, while not having the same functional form as the simplified one used in this work, need to decay in the UV faster than any polynomial in order for the local limit to be recovered.} Nevertheless, this dependence is interesting since it entails a dependence of Huygens' principle violations on the UV structure of the non-local theory and promotes them to potential probes of it.

\section{CONCLUSIONS AND OUTLOOK}\label{secVI}


In this work we have investigated how the signaling between two observers locally coupled to a scalar field is modified by a non-local dynamics for the field. Considering the local interaction of two Unruh--DeWitt detectors with a non-local scalar field, we have shown that whenever communication is allowed in the local theory, the non-locality introduces modifications that are polynomially suppressed by the non-locality scale. 
Furthermore, in the case where the two observers are timelike related --- which does not allow for communication in the local theory --- the signaling is non-vanishing for a non-local field theory, in accordance with the observation that there are SHP violations.

We have explicitly shown that the signaling is exponentially suppressed in the non-locality scale making this effect phenomenologically irrelevant, unless amplification processes are introduced. However, the exponential suppression shows that SHP violations are sensitive to the UV structure of the theory. If this was not the case, we should have obtained a polynomial suppression dictated by the IR behavior. Accordingly, the specific form of the suppression is related to the functional form of the spectral function used. 

The findings of this work lay down the basis for future phenomenological studies of non-locality exploiting Huygens' principle violations as well as other vacuum effects. The non-local corrections to $\mathcal{S}_{2}$, eqs.\eqref{corr1},\eqref{lead}, imply a polynomial in $\ell$ correction to the capacity of the quantum communication channel between Alice and Bob --- defined in terms of bits per use of the channel in~\cite{Jonsson2015}. This opens up the possibility to envisage efficient communication protocols that, together with a high statistic of events, could allow for stringent constraints on the non-locality scale. 

It should be noted that, an analogous polynomial suppression of the response of a single Unruh--DeWitt detector was used in~\cite{PhysRevD.94.061902} to argue for the possibility to cast constraints on the non-locality scale outperforming high-energy experiments. In~\cite{PhysRevD.94.061902}, the effect of non-locality was shown to be more relevant in the case of spontaneous emission, with respect to excitation due to vacuum fluctuations of the field. 

Following the same logic, it would be interesting to consider which initial state of Alice and Bob's detectors maximize the signature of non-locality in a more realistic communication protocol. A simple optimization on initial states is not expected to bring about a different power dependence on the non-locality scale (as one can anticipate from the results in~\cite{Jonsson2}), but a more involved communication protocol may provide an opportunity to accumulate signal and, similar to~\cite{PhysRevD.94.061902}, compensate for the small value of the non-locality scale.

Furthermore, the observation that the polynomial suppression is due to modes of the field localized in a neighborhood of the sender's light-cone gives an indication on to where to look for significant deformations of other effects involving two detectors' interaction with the field's vacuum --- in particular, entanglement harvesting~\cite{Valentini1991,Reznik2003,Pozas2016}.

Finally, in order to arrive at a watertight phenomenological study of non-locality with Unruh--DeWitt detectors, a crucial step is the extension of the model to an Abelian $U(1)$ gauge theory. This extension could be from first principles, like it was done for the scalar field in causal set quantum gravity~\cite{Aslanbeigi:2014zva}, or --- more conservatively --- motivated by the requirement to maintain the same analytic structure for propagators while imposing gauge invariance on the theory. This would result in a non-local electrodynamics and in new effects related to the vectorial nature of non-local fields in such a model. We leave the exploration of this avenue for future works. 


\section*{ACNOWLEDGEMENTS}
AB and DMTB would like to thank Medhi Saravani for helpful discussion concerning the distributional nature of the local limit of the Pauli-Jordan function. DMTB and SL would like to acknowledge financial support from the John Templeton Foundation (JTF), grant  No. 51876. 
AB wish to acknowledge the support of the Austrian Academy of Sciences through Innovationsfonds ''Forschung, Wissenschaft und Gesellschaft``, and the University of Vienna through the research platform TURIS. E.M-M was partially funded by the NSERC Discovery grant programme. E. M-M would like to thank Achim Kempf for his helpful insights.

\appendix
\section{Calculation inside the lightband}\label{appA}
In this appendix we perform the computation of Eq.\eqref{int} (that we rewrite here for convenience)
\begin{align}
S_{2}^{(\ell)}=&-4\alpha_{\textsc{b}}\beta_{\textsc{b}}\alpha_{\textsc{a}}\beta_{\textsc{a}}\int_{\tilde{a}}^{\tilde{b}}dt\cos(t)\int_{t^2 -\tilde{R}^2}^{0}\\ \nonumber
&\frac{d\tilde{y}}{2\sqrt{\tilde{R}^2+\tilde{y}}}\cos(t-\sqrt{\tilde{R}^2+\tilde{y}})\frac{1}{8\pi\tilde{\ell}^2}e^{\frac{-\tilde{y}}{4\tilde{\ell}^2}}.
\end{align}
In order to compute this double integral we proceed as follow. Firstly we compute the inner integral, i.e.,
\begin{align}
&\int_{t^2 -R^2}^{0}\frac{dy}{2\sqrt{R^2+y}}\cos(t-\sqrt{R^2+y})\frac{1}{8\pi\ell^2}e^{\frac{-y}{4\ell^2}}\\ \nonumber
&= \frac{i e^{\frac{R^2}{4 \ell^2}-\ell^2} \left(2 \ii \Im\left(e^{\ii t} \text{erfi}\left(\ell-\frac{\ii R}{2 \ell}\right)\right)+2 \ii \Im\left(e^{-\ii t} \text{erfi}\left(\ell+\frac{\ii t}{2 \ell}\right)\right)\right)}{16 \sqrt{\pi } \ell}
\end{align}
It should be noted that, in accordance with the discussion in the main text about the distributional local limit of the Pauli-Jordan function, in the local limit this expression reduces to $-\cos (R-t)/4 \pi  R$, i.e., to the result of 
\begin{align}
&\int_{t^2 -R^2}^{0}\frac{dy}{2\sqrt{R^2+y}}\cos(t-\sqrt{R^2+y})\left(-\frac{\delta(y)}{2\pi}\right).
\end{align}

Now, the remaining integral can be performed giving the final result
\begin{widetext}
\begin{align}\label{gen1}
    S_{2}^{(\ell)}=& -4\alpha_{\textsc{b}}\beta_{\textsc{b}}\alpha_{\textsc{a}}\beta_{\textsc{a}}\\ \nonumber
    &\frac{e^{\frac{R^2}{4 l^2}}}{64 \sqrt{\pi } \ell} \left[\frac{8 \ell e^{-\frac{a^2}{4 l^2}} \cos (a)}{\sqrt{\pi }}+e^{-\ell^2} \left(2 \Re\left(\left(e^{2 i b}-e^{2 i a}\right) \text{erfi}\left(\ell-\frac{\ii R}{2 \ell}\right)\right)-4 (a-b) \Im\left(\text{erfi}\left(\ell+\frac{\ii R}{2 \ell}\right)\right)\right)\right.\\ \nonumber
    &\left.+e^{-\ell^2} \left(\left(8 \ell^2-2\right) \Im\left(\text{erf}\left(\frac{a}{2 \ell}-\ii \ell\right)\right)-2 \Im\left(e^{-2 i a} \text{erf}\left(\frac{a}{2 \ell}-\ii \ell\right)\right)+4 a \Re\left(\text{erf}\left(\frac{a}{2 \ell}-\ii \ell\right)\right)\right)\right.\\ \nonumber
    &\left.-\frac{8 l e^{-\frac{b^2}{4 l^2}} \cos (b)}{\sqrt{\pi }}+2 e^{-l^2} \left(\left(1-4 l^2\right) \Im\left(\text{erf}\left(\frac{b}{2 \ell}-\ii \ell\right)\right)+\Im\left(e^{2 \ii b} \text{erf}\left(\frac{b}{2 \ell}-\ii \ell\right)\right)\right.\right.\\ \nonumber
    &\left.\left. -2 b \Re\left(\text{erf}\left(\frac{b}{2 \ell}-\ii \ell\right)\right)\right)\right].
\end{align}
\end{widetext}
When specializing to the case $a=R,\; b=R+T$, it is possible to take the limit $\lim_{\ell\rightarrow 0^+}\left(S_{2}^{local}+S_{2}^{(\ell)}\right)/\ell^2$ which, in turn, gives the leading correction in eq.\eqref{lead}.

\bibliography{references2}

\begin{thebibliography}{37}%
\makeatletter
\providecommand \@ifxundefined [1]{%
 \@ifx{#1\undefined}
}%
\providecommand \@ifnum [1]{%
 \ifnum #1\expandafter \@firstoftwo
 \else \expandafter \@secondoftwo
 \fi
}%
\providecommand \@ifx [1]{%
 \ifx #1\expandafter \@firstoftwo
 \else \expandafter \@secondoftwo
 \fi
}%
\providecommand \natexlab [1]{#1}%
\providecommand \enquote  [1]{``#1''}%
\providecommand \bibnamefont  [1]{#1}%
\providecommand \bibfnamefont [1]{#1}%
\providecommand \citenamefont [1]{#1}%
\providecommand \href@noop [0]{\@secondoftwo}%
\providecommand \href [0]{\begingroup \@sanitize@url \@href}%
\providecommand \@href[1]{\@@startlink{#1}\@@href}%
\providecommand \@@href[1]{\endgroup#1\@@endlink}%
\providecommand \@sanitize@url [0]{\catcode `\\12\catcode `\$12\catcode
  `\&12\catcode `\#12\catcode `\^12\catcode `\_12\catcode `\%12\relax}%
\providecommand \@@startlink[1]{}%
\providecommand \@@endlink[0]{}%
\providecommand \url  [0]{\begingroup\@sanitize@url \@url }%
\providecommand \@url [1]{\endgroup\@href {#1}{\urlprefix }}%
\providecommand \urlprefix  [0]{URL }%
\providecommand \Eprint [0]{\href }%
\providecommand \doibase [0]{http://dx.doi.org/}%
\providecommand \selectlanguage [0]{\@gobble}%
\providecommand \bibinfo  [0]{\@secondoftwo}%
\providecommand \bibfield  [0]{\@secondoftwo}%
\providecommand \translation [1]{[#1]}%
\providecommand \BibitemOpen [0]{}%
\providecommand \bibitemStop [0]{}%
\providecommand \bibitemNoStop [0]{.\EOS\space}%
\providecommand \EOS [0]{\spacefactor3000\relax}%
\providecommand \BibitemShut  [1]{\csname bibitem#1\endcsname}%
\let\auto@bib@innerbib\@empty
\bibitem [{\citenamefont {Ellis}\ and\ \citenamefont {Sciama}(1972)}]{Ellis}%
  \BibitemOpen
  \bibfield  {author} {\bibinfo {author} {\bibfnamefont {G.~F.~R.}\
  \bibnamefont {Ellis}}\ and\ \bibinfo {author} {\bibfnamefont {D.~W.}\
  \bibnamefont {Sciama}},\ }\href@noop {} {\emph {\bibinfo {title} {in L.
  O'Raifeartaigh, ed., General Relativity, Papers in Honour of J. L. Synge}}}\
  (\bibinfo  {publisher} {Oxford: Clarendon Press},\ \bibinfo {year}
  {1972})\BibitemShut {NoStop}%
\bibitem [{\citenamefont {McLenaghan}(1974)}]{McLenaghan}%
  \BibitemOpen
  \bibfield  {author} {\bibinfo {author} {\bibfnamefont {R.}~\bibnamefont
  {McLenaghan}},\ }\href@noop {} {\bibfield  {journal} {\bibinfo  {journal}
  {Ann. Inst. H. Poincare}\ }\textbf {\bibinfo {volume} {20}},\ \bibinfo
  {pages} {153} (\bibinfo {year} {1974})}\BibitemShut {NoStop}%
\bibitem [{\citenamefont {Sonego}\ and\ \citenamefont
  {Faraoni}(1992)}]{Sonego:1991sq}%
  \BibitemOpen
  \bibfield  {author} {\bibinfo {author} {\bibfnamefont {S.}~\bibnamefont
  {Sonego}}\ and\ \bibinfo {author} {\bibfnamefont {V.}~\bibnamefont
  {Faraoni}},\ }\href {\doibase 10.1063/1.529798} {\bibfield  {journal}
  {\bibinfo  {journal} {J. Math. Phys.}\ }\textbf {\bibinfo {volume} {33}},\
  \bibinfo {pages} {625} (\bibinfo {year} {1992})}\BibitemShut {NoStop}%
\bibitem [{\citenamefont {Czapor}\ and\ \citenamefont
  {McLenaghan}(2008)}]{czapor}%
  \BibitemOpen
  \bibfield  {author} {\bibinfo {author} {\bibfnamefont {S.}~\bibnamefont
  {Czapor}}\ and\ \bibinfo {author} {\bibfnamefont {R.}~\bibnamefont
  {McLenaghan}},\ }\href
  {http://www.actaphys.uj.edu.pl/_old/sup1/pdf/s1p0055.pdf} {\bibfield
  {journal} {\bibinfo  {journal} {Acta. Phys. Pol. B Proc. Suppl. 1}\ }\textbf
  {\bibinfo {volume} {1}},\ \bibinfo {pages} {55} (\bibinfo {year}
  {2008})}\BibitemShut {NoStop}%
\bibitem [{\citenamefont {Jonsson}\ \emph {et~al.}(2015)\citenamefont
  {Jonsson}, \citenamefont {Mart\'{i}n-Mart\'{i}nez},\ and\ \citenamefont
  {Kempf}}]{Jonsson2015}%
  \BibitemOpen
  \bibfield  {author} {\bibinfo {author} {\bibfnamefont {R.~H.}\ \bibnamefont
  {Jonsson}}, \bibinfo {author} {\bibfnamefont {E.}~\bibnamefont
  {Mart\'{i}n-Mart\'{i}nez}}, \ and\ \bibinfo {author} {\bibfnamefont
  {A.}~\bibnamefont {Kempf}},\ }\href {\doibase 10.1103/PhysRevLett.114.110505}
  {\bibfield  {journal} {\bibinfo  {journal} {Phys. Rev. Lett.}\ }\textbf
  {\bibinfo {volume} {114}},\ \bibinfo {pages} {110505} (\bibinfo {year}
  {2015})}\BibitemShut {NoStop}%
\bibitem [{\citenamefont {Belenchia}\ \emph
  {et~al.}(2015{\natexlab{a}})\citenamefont {Belenchia}, \citenamefont
  {Benincasa},\ and\ \citenamefont {Liberati}}]{Belenchia:2014fda}%
  \BibitemOpen
  \bibfield  {author} {\bibinfo {author} {\bibfnamefont {A.}~\bibnamefont
  {Belenchia}}, \bibinfo {author} {\bibfnamefont {D.~M.~T.}\ \bibnamefont
  {Benincasa}}, \ and\ \bibinfo {author} {\bibfnamefont {S.}~\bibnamefont
  {Liberati}},\ }\href {\doibase 10.1007/JHEP03(2015)036} {\bibfield  {journal}
  {\bibinfo  {journal} {JHEP}\ }\textbf {\bibinfo {volume} {03}},\ \bibinfo
  {pages} {036} (\bibinfo {year} {2015}{\natexlab{a}})},\ \Eprint
  {http://arxiv.org/abs/1411.6513} {arXiv:1411.6513 [gr-qc]} \BibitemShut
  {NoStop}%
\bibitem [{\citenamefont {Bombelli}\ \emph {et~al.}(1987)\citenamefont
  {Bombelli}, \citenamefont {Lee}, \citenamefont {Meyer},\ and\ \citenamefont
  {Sorkin}}]{PhysRevLett.59.521}%
  \BibitemOpen
  \bibfield  {author} {\bibinfo {author} {\bibfnamefont {L.}~\bibnamefont
  {Bombelli}}, \bibinfo {author} {\bibfnamefont {J.}~\bibnamefont {Lee}},
  \bibinfo {author} {\bibfnamefont {D.}~\bibnamefont {Meyer}}, \ and\ \bibinfo
  {author} {\bibfnamefont {R.~D.}\ \bibnamefont {Sorkin}},\ }\href {\doibase
  10.1103/PhysRevLett.59.521} {\bibfield  {journal} {\bibinfo  {journal} {Phys.
  Rev. Lett.}\ }\textbf {\bibinfo {volume} {59}},\ \bibinfo {pages} {521}
  (\bibinfo {year} {1987})}\BibitemShut {NoStop}%
\bibitem [{\citenamefont {Sorkin}(2006)}]{Sorkin2006}%
  \BibitemOpen
  \bibfield  {author} {\bibinfo {author} {\bibfnamefont {R.~D.}\ \bibnamefont
  {Sorkin}},\ }in\ \href@noop {} {\emph {\bibinfo {booktitle} {{Approaches to
  Quantum Gravity: Towards a New Understanding of Space and Time}}}},\ \bibinfo
  {editor} {edited by\ \bibinfo {editor} {\bibfnamefont {D.}~\bibnamefont
  {Oriti}}}\ (\bibinfo  {publisher} {Cambridge University Press},\ \bibinfo
  {year} {2006})\ \Eprint {http://arxiv.org/abs/gr-qc/0703099}
  {arXiv:gr-qc/0703099} \BibitemShut {NoStop}%
\bibitem [{\citenamefont {Belenchia}\ \emph
  {et~al.}(2015{\natexlab{b}})\citenamefont {Belenchia}, \citenamefont
  {Benincasa},\ and\ \citenamefont {Dowker}}]{Belenchia:2015hca}%
  \BibitemOpen
  \bibfield  {author} {\bibinfo {author} {\bibfnamefont {A.}~\bibnamefont
  {Belenchia}}, \bibinfo {author} {\bibfnamefont {D.~M.}\ \bibnamefont
  {Benincasa}}, \ and\ \bibinfo {author} {\bibfnamefont {F.}~\bibnamefont
  {Dowker}},\ }\href@noop {} {\  (\bibinfo {year}
  {2015}{\natexlab{b}})}\BibitemShut {NoStop}%
\bibitem [{\citenamefont {Benincasa}\ and\ \citenamefont
  {Dowker}(2010)}]{Benincasa:2010ac}%
  \BibitemOpen
  \bibfield  {author} {\bibinfo {author} {\bibfnamefont {D.~M.~T.}\
  \bibnamefont {Benincasa}}\ and\ \bibinfo {author} {\bibfnamefont
  {F.}~\bibnamefont {Dowker}},\ }\href {\doibase
  10.1103/PhysRevLett.104.181301} {\bibfield  {journal} {\bibinfo  {journal}
  {Phys. Rev. Lett.}\ }\textbf {\bibinfo {volume} {104}},\ \bibinfo {pages}
  {181301} (\bibinfo {year} {2010})},\ \Eprint {http://arxiv.org/abs/1001.2725}
  {arXiv:1001.2725 [Unknown]} \BibitemShut {NoStop}%
\bibitem [{\citenamefont {Dowker}\ and\ \citenamefont
  {Glaser}(2013)}]{Dowker:2013vba}%
  \BibitemOpen
  \bibfield  {author} {\bibinfo {author} {\bibfnamefont {F.}~\bibnamefont
  {Dowker}}\ and\ \bibinfo {author} {\bibfnamefont {L.}~\bibnamefont
  {Glaser}},\ }\href {\doibase 10.1088/0264-9381/30/19/195016} {\bibfield
  {journal} {\bibinfo  {journal} {Class. Quant. Grav.}\ }\textbf {\bibinfo
  {volume} {30}},\ \bibinfo {pages} {195016} (\bibinfo {year} {2013})},\
  \Eprint {http://arxiv.org/abs/1305.2588} {arXiv:1305.2588 [gr-qc]}
  \BibitemShut {NoStop}%
\bibitem [{\citenamefont {Glaser}(2013)}]{Glaser:2013sf}%
  \BibitemOpen
  \bibfield  {author} {\bibinfo {author} {\bibfnamefont {L.}~\bibnamefont
  {Glaser}},\ }\href {http://arxiv.org/abs/1311.1701} {\  (\bibinfo {year}
  {2013})},\ \Eprint {http://arxiv.org/abs/1311.1701} {1311.1701} \BibitemShut
  {NoStop}%
\bibitem [{\citenamefont {Belenchia}(2016)}]{BelenchiaUni}%
  \BibitemOpen
  \bibfield  {author} {\bibinfo {author} {\bibfnamefont {A.}~\bibnamefont
  {Belenchia}},\ }\href {http://stacks.iop.org/0264-9381/33/i=13/a=135011}
  {\bibfield  {journal} {\bibinfo  {journal} {Classical and Quantum Gravity}\
  }\textbf {\bibinfo {volume} {33}},\ \bibinfo {pages} {135011} (\bibinfo
  {year} {2016})}\BibitemShut {NoStop}%
\bibitem [{\citenamefont {Belenchia}\ \emph {et~al.}(2016)\citenamefont
  {Belenchia}, \citenamefont {Benincasa}, \citenamefont
  {Mart\'{\i}n-Mart\'{\i}nez},\ and\ \citenamefont
  {Saravani}}]{PhysRevD.94.061902}%
  \BibitemOpen
  \bibfield  {author} {\bibinfo {author} {\bibfnamefont {A.}~\bibnamefont
  {Belenchia}}, \bibinfo {author} {\bibfnamefont {D.~M.~T.}\ \bibnamefont
  {Benincasa}}, \bibinfo {author} {\bibfnamefont {E.}~\bibnamefont
  {Mart\'{\i}n-Mart\'{\i}nez}}, \ and\ \bibinfo {author} {\bibfnamefont
  {M.}~\bibnamefont {Saravani}},\ }\href {\doibase 10.1103/PhysRevD.94.061902}
  {\bibfield  {journal} {\bibinfo  {journal} {Phys. Rev. D}\ }\textbf {\bibinfo
  {volume} {94}},\ \bibinfo {pages} {061902} (\bibinfo {year}
  {2016})}\BibitemShut {NoStop}%
\bibitem [{\citenamefont {DeWitt}(1980)}]{DeWitt1980}%
  \BibitemOpen
  \bibfield  {author} {\bibinfo {author} {\bibfnamefont {B.}~\bibnamefont
  {DeWitt}},\ }\href@noop {} {\emph {\bibinfo {title} {General Relativity; an
  Einstein Centenary Survey}}}\ (\bibinfo  {publisher} {Cambridge University
  Press},\ \bibinfo {year} {1980})\BibitemShut {NoStop}%
\bibitem [{\citenamefont {Mart\'{i}n-Mart\'{i}nez}\ \emph
  {et~al.}(2013)\citenamefont {Mart\'{i}n-Mart\'{i}nez}, \citenamefont
  {Montero},\ and\ \citenamefont {del Rey}}]{Martin-Martinez2013}%
  \BibitemOpen
  \bibfield  {author} {\bibinfo {author} {\bibfnamefont {E.}~\bibnamefont
  {Mart\'{i}n-Mart\'{i}nez}}, \bibinfo {author} {\bibfnamefont
  {M.}~\bibnamefont {Montero}}, \ and\ \bibinfo {author} {\bibfnamefont
  {M.}~\bibnamefont {del Rey}},\ }\href {\doibase 10.1103/PhysRevD.87.064038}
  {\bibfield  {journal} {\bibinfo  {journal} {Phys. Rev. D}\ }\textbf {\bibinfo
  {volume} {87}},\ \bibinfo {pages} {064038} (\bibinfo {year}
  {2013})}\BibitemShut {NoStop}%
\bibitem [{\citenamefont {Alhambra}\ \emph {et~al.}(2014)\citenamefont
  {Alhambra}, \citenamefont {Kempf},\ and\ \citenamefont
  {Mart\'{\i}n-Mart\'{\i}nez}}]{Alhambra2014}%
  \BibitemOpen
  \bibfield  {author} {\bibinfo {author} {\bibfnamefont {A.~M.}\ \bibnamefont
  {Alhambra}}, \bibinfo {author} {\bibfnamefont {A.}~\bibnamefont {Kempf}}, \
  and\ \bibinfo {author} {\bibfnamefont {E.}~\bibnamefont
  {Mart\'{\i}n-Mart\'{\i}nez}},\ }\href {\doibase 10.1103/PhysRevA.89.033835}
  {\bibfield  {journal} {\bibinfo  {journal} {Phys. Rev. A}\ }\textbf {\bibinfo
  {volume} {89}},\ \bibinfo {pages} {033835} (\bibinfo {year}
  {2014})}\BibitemShut {NoStop}%
\bibitem [{\citenamefont {Pozas-Kerstjens}\ and\ \citenamefont
  {Mart\'{\i}n-Mart\'{\i}nez}(2016)}]{Pozas2016}%
  \BibitemOpen
  \bibfield  {author} {\bibinfo {author} {\bibfnamefont {A.}~\bibnamefont
  {Pozas-Kerstjens}}\ and\ \bibinfo {author} {\bibfnamefont {E.}~\bibnamefont
  {Mart\'{\i}n-Mart\'{\i}nez}},\ }\href {\doibase 10.1103/PhysRevD.94.064074}
  {\bibfield  {journal} {\bibinfo  {journal} {Phys. Rev. D}\ }\textbf {\bibinfo
  {volume} {94}},\ \bibinfo {pages} {064074} (\bibinfo {year}
  {2016})}\BibitemShut {NoStop}%
\bibitem [{\citenamefont {Jonsson}\ \emph {et~al.}(2014)\citenamefont
  {Jonsson}, \citenamefont {Mart\'{\i}n-Mart\'{\i}nez},\ and\ \citenamefont
  {Kempf}}]{Jonsson2014}%
  \BibitemOpen
  \bibfield  {author} {\bibinfo {author} {\bibfnamefont {R.~H.}\ \bibnamefont
  {Jonsson}}, \bibinfo {author} {\bibfnamefont {E.}~\bibnamefont
  {Mart\'{\i}n-Mart\'{\i}nez}}, \ and\ \bibinfo {author} {\bibfnamefont
  {A.}~\bibnamefont {Kempf}},\ }\href {\doibase 10.1103/PhysRevA.89.022330}
  {\bibfield  {journal} {\bibinfo  {journal} {Phys. Rev. A}\ }\textbf {\bibinfo
  {volume} {89}},\ \bibinfo {pages} {022330} (\bibinfo {year}
  {2014})}\BibitemShut {NoStop}%
\bibitem [{\citenamefont {Blasco}\ \emph
  {et~al.}(2015{\natexlab{a}})\citenamefont {Blasco}, \citenamefont {Garay},
  \citenamefont {Mart\'{i}n-Benito},\ and\ \citenamefont
  {Mart\'{\i}n-Mart\'{\i}nez}}]{Bounce2}%
  \BibitemOpen
  \bibfield  {author} {\bibinfo {author} {\bibfnamefont {A.}~\bibnamefont
  {Blasco}}, \bibinfo {author} {\bibfnamefont {L.~J.}\ \bibnamefont {Garay}},
  \bibinfo {author} {\bibfnamefont {M.}~\bibnamefont {Mart\'{i}n-Benito}}, \
  and\ \bibinfo {author} {\bibfnamefont {E.}~\bibnamefont
  {Mart\'{\i}n-Mart\'{\i}nez}},\ }\href {\doibase 10.1139/cjp-2014-0567}
  {\bibfield  {journal} {\bibinfo  {journal} {Canadian Journal of Physics}\
  }\textbf {\bibinfo {volume} {93}},\ \bibinfo {pages} {968} (\bibinfo {year}
  {2015}{\natexlab{a}})}\BibitemShut {NoStop}%
\bibitem [{\citenamefont {Garay}\ \emph {et~al.}(2014)\citenamefont {Garay},
  \citenamefont {Mart\'{\i}n-Benito},\ and\ \citenamefont
  {Mart\'{\i}n-Mart\'{\i}nez}}]{Bounce12}%
  \BibitemOpen
  \bibfield  {author} {\bibinfo {author} {\bibfnamefont {L.~J.}\ \bibnamefont
  {Garay}}, \bibinfo {author} {\bibfnamefont {M.}~\bibnamefont
  {Mart\'{\i}n-Benito}}, \ and\ \bibinfo {author} {\bibfnamefont
  {E.}~\bibnamefont {Mart\'{\i}n-Mart\'{\i}nez}},\ }\href {\doibase
  10.1103/PhysRevD.89.043510} {\bibfield  {journal} {\bibinfo  {journal} {Phys.
  Rev. D}\ }\textbf {\bibinfo {volume} {89}},\ \bibinfo {pages} {043510}
  (\bibinfo {year} {2014})}\BibitemShut {NoStop}%
\bibitem [{\citenamefont {Blasco}\ \emph
  {et~al.}(2016{\natexlab{a}})\citenamefont {Blasco}, \citenamefont {Garay},
  \citenamefont {Mart\'{\i}n-Benito},\ and\ \citenamefont
  {Mart\'{\i}n-Mart\'{\i}nez}}]{Blascoprd}%
  \BibitemOpen
  \bibfield  {author} {\bibinfo {author} {\bibfnamefont {A.}~\bibnamefont
  {Blasco}}, \bibinfo {author} {\bibfnamefont {L.~J.}\ \bibnamefont {Garay}},
  \bibinfo {author} {\bibfnamefont {M.}~\bibnamefont {Mart\'{\i}n-Benito}}, \
  and\ \bibinfo {author} {\bibfnamefont {E.}~\bibnamefont
  {Mart\'{\i}n-Mart\'{\i}nez}},\ }\href {\doibase 10.1103/PhysRevD.93.024055}
  {\bibfield  {journal} {\bibinfo  {journal} {Phys. Rev. D}\ }\textbf {\bibinfo
  {volume} {93}},\ \bibinfo {pages} {024055} (\bibinfo {year}
  {2016}{\natexlab{a}})}\BibitemShut {NoStop}%
\bibitem [{\citenamefont {Blasco}\ \emph
  {et~al.}(2015{\natexlab{b}})\citenamefont {Blasco}, \citenamefont {Garay},
  \citenamefont {Martin-Benito},\ and\ \citenamefont
  {Mart\'{i}n-Mart\'{i}nez}}]{Blasco:2015eya}%
  \BibitemOpen
  \bibfield  {author} {\bibinfo {author} {\bibfnamefont {A.}~\bibnamefont
  {Blasco}}, \bibinfo {author} {\bibfnamefont {L.~J.}\ \bibnamefont {Garay}},
  \bibinfo {author} {\bibfnamefont {M.}~\bibnamefont {Martin-Benito}}, \ and\
  \bibinfo {author} {\bibfnamefont {E.}~\bibnamefont
  {Mart\'{i}n-Mart\'{i}nez}},\ }\href {\doibase 10.1103/PhysRevLett.114.141103}
  {\bibfield  {journal} {\bibinfo  {journal} {Phys. Rev. Lett.}\ }\textbf
  {\bibinfo {volume} {114}},\ \bibinfo {pages} {141103} (\bibinfo {year}
  {2015}{\natexlab{b}})},\ \Eprint {http://arxiv.org/abs/1501.01650}
  {arXiv:1501.01650 [quant-ph]} \BibitemShut {NoStop}%
\bibitem [{\citenamefont
  {Mart\'{\i}n-Mart\'{\i}nez}(2015)}]{Martin-Martinez2015}%
  \BibitemOpen
  \bibfield  {author} {\bibinfo {author} {\bibfnamefont {E.}~\bibnamefont
  {Mart\'{\i}n-Mart\'{\i}nez}},\ }\href {\doibase 10.1103/PhysRevD.92.104019}
  {\bibfield  {journal} {\bibinfo  {journal} {Phys. Rev. D}\ }\textbf {\bibinfo
  {volume} {92}},\ \bibinfo {pages} {104019} (\bibinfo {year}
  {2015})}\BibitemShut {NoStop}%
\bibitem [{\citenamefont {Landulfo}(2016)}]{Landulfo2016}%
  \BibitemOpen
  \bibfield  {author} {\bibinfo {author} {\bibfnamefont {A.~G.~S.}\
  \bibnamefont {Landulfo}},\ }\href {\doibase 10.1103/PhysRevD.93.104019}
  {\bibfield  {journal} {\bibinfo  {journal} {Phys. Rev. D}\ }\textbf {\bibinfo
  {volume} {93}},\ \bibinfo {pages} {104019} (\bibinfo {year}
  {2016})}\BibitemShut {NoStop}%
\bibitem [{\citenamefont {Simidzija}\ and\ \citenamefont
  {Mart\'{\i}n-Mart\'{\i}nez}(2017{\natexlab{a}})}]{Petar}%
  \BibitemOpen
  \bibfield  {author} {\bibinfo {author} {\bibfnamefont {P.}~\bibnamefont
  {Simidzija}}\ and\ \bibinfo {author} {\bibfnamefont {E.}~\bibnamefont
  {Mart\'{\i}n-Mart\'{\i}nez}},\ }\href {\doibase 10.1103/PhysRevD.95.025002}
  {\bibfield  {journal} {\bibinfo  {journal} {Phys. Rev. D}\ }\textbf {\bibinfo
  {volume} {95}},\ \bibinfo {pages} {025002} (\bibinfo {year}
  {2017}{\natexlab{a}})}\BibitemShut {NoStop}%
\bibitem [{\citenamefont {Jonsson}(2016{\natexlab{a}})}]{Jonsson1}%
  \BibitemOpen
  \bibfield  {author} {\bibinfo {author} {\bibfnamefont {R.~H.}\ \bibnamefont
  {Jonsson}},\ }\href {http://stacks.iop.org/1751-8121/49/i=44/a=445402}
  {\bibfield  {journal} {\bibinfo  {journal} {Journal of Physics A:
  Mathematical and Theoretical}\ }\textbf {\bibinfo {volume} {49}},\ \bibinfo
  {pages} {445402} (\bibinfo {year} {2016}{\natexlab{a}})}\BibitemShut
  {NoStop}%
\bibitem [{\citenamefont {Jonsson}(2017)}]{Jonsson2}%
  \BibitemOpen
  \bibfield  {author} {\bibinfo {author} {\bibfnamefont {R.~H.}\ \bibnamefont
  {Jonsson}},\ }\href@noop {} {\enquote {\bibinfo {title} {Quantum signaling in
  relativistic motion and across acceleration horizons},}\ } (\bibinfo {year}
  {2017}),\ \Eprint {http://arxiv.org/abs/arXiv:1702.06847} {arXiv:1702.06847}
  \BibitemShut {NoStop}%
\bibitem [{\citenamefont {Simidzija}\ and\ \citenamefont
  {Mart\'{\i}n-Mart\'{\i}nez}(2017{\natexlab{b}})}]{Petar2017}%
  \BibitemOpen
  \bibfield  {author} {\bibinfo {author} {\bibfnamefont {P.}~\bibnamefont
  {Simidzija}}\ and\ \bibinfo {author} {\bibfnamefont {E.}~\bibnamefont
  {Mart\'{\i}n-Mart\'{\i}nez}},\ }\href {\doibase 10.1103/PhysRevD.95.025002}
  {\bibfield  {journal} {\bibinfo  {journal} {Phys. Rev. D}\ }\textbf {\bibinfo
  {volume} {95}},\ \bibinfo {pages} {025002} (\bibinfo {year}
  {2017}{\natexlab{b}})}\BibitemShut {NoStop}%
\bibitem [{\citenamefont {Blasco}\ \emph
  {et~al.}(2016{\natexlab{b}})\citenamefont {Blasco}, \citenamefont {Garay},
  \citenamefont {Mart\'{\i}n-Benito},\ and\ \citenamefont
  {Mart\'{\i}n-Mart\'{\i}nez}}]{PhysRevD.93.024055}%
  \BibitemOpen
  \bibfield  {author} {\bibinfo {author} {\bibfnamefont {A.}~\bibnamefont
  {Blasco}}, \bibinfo {author} {\bibfnamefont {L.~J.}\ \bibnamefont {Garay}},
  \bibinfo {author} {\bibfnamefont {M.}~\bibnamefont {Mart\'{\i}n-Benito}}, \
  and\ \bibinfo {author} {\bibfnamefont {E.}~\bibnamefont
  {Mart\'{\i}n-Mart\'{\i}nez}},\ }\href {\doibase 10.1103/PhysRevD.93.024055}
  {\bibfield  {journal} {\bibinfo  {journal} {Phys. Rev. D}\ }\textbf {\bibinfo
  {volume} {93}},\ \bibinfo {pages} {024055} (\bibinfo {year}
  {2016}{\natexlab{b}})}\BibitemShut {NoStop}%
\bibitem [{\citenamefont {Jonsson}(2016{\natexlab{b}})}]{Robert1}%
  \BibitemOpen
  \bibfield  {author} {\bibinfo {author} {\bibfnamefont {R.~H.}\ \bibnamefont
  {Jonsson}},\ }\href {http://stacks.iop.org/1751-8121/49/i=44/a=445402}
  {\bibfield  {journal} {\bibinfo  {journal} {Journal of Physics A:
  Mathematical and Theoretical}\ }\textbf {\bibinfo {volume} {49}},\ \bibinfo
  {pages} {445402} (\bibinfo {year} {2016}{\natexlab{b}})}\BibitemShut
  {NoStop}%
\bibitem [{\citenamefont {Aslanbeigi}\ \emph {et~al.}(2014)\citenamefont
  {Aslanbeigi}, \citenamefont {Saravani},\ and\ \citenamefont
  {Sorkin}}]{Aslanbeigi:2014zva}%
  \BibitemOpen
  \bibfield  {author} {\bibinfo {author} {\bibfnamefont {S.}~\bibnamefont
  {Aslanbeigi}}, \bibinfo {author} {\bibfnamefont {M.}~\bibnamefont
  {Saravani}}, \ and\ \bibinfo {author} {\bibfnamefont {R.~D.}\ \bibnamefont
  {Sorkin}},\ }\href {\doibase 10.1007/JHEP06(2014)024} {\bibfield  {journal}
  {\bibinfo  {journal} {JHEP}\ }\textbf {\bibinfo {volume} {1406}},\ \bibinfo
  {pages} {024} (\bibinfo {year} {2014})},\ \Eprint
  {http://arxiv.org/abs/1403.1622} {arXiv:1403.1622 [hep-th]} \BibitemShut
  {NoStop}%
\bibitem [{\citenamefont {Saravani}\ and\ \citenamefont
  {Aslanbeigi}(2015{\natexlab{a}})}]{Saravani:2015rva}%
  \BibitemOpen
  \bibfield  {author} {\bibinfo {author} {\bibfnamefont {M.}~\bibnamefont
  {Saravani}}\ and\ \bibinfo {author} {\bibfnamefont {S.}~\bibnamefont
  {Aslanbeigi}},\ }\href {\doibase 10.1103/PhysRevD.92.103504} {\bibfield
  {journal} {\bibinfo  {journal} {Phys. Rev. D}\ }\textbf {\bibinfo {volume}
  {92}},\ \bibinfo {pages} {103504} (\bibinfo {year}
  {2015}{\natexlab{a}})}\BibitemShut {NoStop}%
\bibitem [{\citenamefont {Saravani}\ and\ \citenamefont
  {Aslanbeigi}(2015{\natexlab{b}})}]{PhysRevD.92.103504}%
  \BibitemOpen
  \bibfield  {author} {\bibinfo {author} {\bibfnamefont {M.}~\bibnamefont
  {Saravani}}\ and\ \bibinfo {author} {\bibfnamefont {S.}~\bibnamefont
  {Aslanbeigi}},\ }\href {\doibase 10.1103/PhysRevD.92.103504} {\bibfield
  {journal} {\bibinfo  {journal} {Phys. Rev. D}\ }\textbf {\bibinfo {volume}
  {92}},\ \bibinfo {pages} {103504} (\bibinfo {year}
  {2015}{\natexlab{b}})}\BibitemShut {NoStop}%
\bibitem [{\citenamefont {Bogoliubov}\ \emph {et~al.}(1959)\citenamefont
  {Bogoliubov}, \citenamefont {Shirkov},\ and\ \citenamefont
  {Chomet}}]{bogoliubov1959introduction}%
  \BibitemOpen
  \bibfield  {author} {\bibinfo {author} {\bibfnamefont {N.~N.}\ \bibnamefont
  {Bogoliubov}}, \bibinfo {author} {\bibfnamefont {D.~V.}\ \bibnamefont
  {Shirkov}}, \ and\ \bibinfo {author} {\bibfnamefont {S.}~\bibnamefont
  {Chomet}},\ }\href@noop {} {\emph {\bibinfo {title} {Introduction to the
  theory of quantized fields}}},\ Vol.~\bibinfo {volume} {59}\ (\bibinfo
  {publisher} {Interscience New York},\ \bibinfo {year} {1959})\BibitemShut
  {NoStop}%
\bibitem [{\citenamefont {Valentini}(1991)}]{Valentini1991}%
  \BibitemOpen
  \bibfield  {author} {\bibinfo {author} {\bibfnamefont {A.}~\bibnamefont
  {Valentini}},\ }\href {\doibase
  http://dx.doi.org/10.1016/0375-9601(91)90952-5} {\bibfield  {journal}
  {\bibinfo  {journal} {Physics Letters A}\ }\textbf {\bibinfo {volume}
  {153}},\ \bibinfo {pages} {321 } (\bibinfo {year} {1991})}\BibitemShut
  {NoStop}%
\bibitem [{\citenamefont {Reznik}(2003)}]{Reznik2003}%
  \BibitemOpen
  \bibfield  {author} {\bibinfo {author} {\bibfnamefont {B.}~\bibnamefont
  {Reznik}},\ }\href {\doibase 10.1023/A:1022875910744} {\bibfield  {journal}
  {\bibinfo  {journal} {Foundations of Physics}\ }\textbf {\bibinfo {volume}
  {33}},\ \bibinfo {pages} {167} (\bibinfo {year} {2003})}\BibitemShut
  {NoStop}%
\end{thebibliography}%

\end{document}